\documentclass[12pt]{article}
\usepackage{amssymb}
\usepackage{amsmath}
\usepackage{theorem}
\usepackage{latexsym}
\usepackage{amscd}
\theoremheaderfont{\bf} %\theoremstyle{break}
\theorembodyfont{\sl}
\newtheorem{theo}{Theorem}[section]
{\theorembodyfont{\rm} \newtheorem{defi}[theo]{Definition}}
{\theorembodyfont{\rm} \newtheorem{exa}[theo]{Example}}
{\theorembodyfont{\rm} \newtheorem{rem}[theo]{Remark}}
{\theorembodyfont{\rm} }
{\theorembodyfont{\rm} }

\newtheorem{cor}[theo]{Corollary}
\newtheorem{lemma}[theo]{Lemma}
\newtheorem{conjecture}[theo]{Conjecture}

\newenvironment{proof}{{\sc Proof:}}{\mbox{}\hfill$\Box$\par}
\newcommand{\junk}[1]{}
\newcommand{\N}{{\mathbb N}}
\newcommand{\F}{{\mathbb F}}

\newcommand{\Z}{{\mathbb Z}}
\newcommand{\C}{{\mathcal C}}

\newcommand{\wt}{{\rm wt}}

\newcounter{abc}

\title{Superregular Matrices and the Construction of Convolutional Codes having
a Maximum Distance Profile}
       
\author{
  Ryan Hutchinson\\
  {\small Mathematics Institute}\vspace{-2mm}\\
  {\small University of Z\"urich}\vspace{-2mm}\\
  {\small Z\"urich, Switzerland}\vspace{-2mm}\\
  {\small {\em e-mail:} rhutchin@math.unizh.ch}
  \and
  Roxana Smarandache\thanks{On leave at the University of Notre Dame, Department of Mathematics, Notre Dame, IN 46556.} \\
  {\small Department of Mathematics and Statistics}\vspace{-2mm}\\
  {\small San Diego State University}\vspace{-2mm}\\
  {\small San Diego, CA 92182-7720, USA}\vspace{-2mm}\\
  {\small {\em e-mail:} rsmarand@sciences.sdsu.edu }
  \and
  Jochen Trumpf\thanks{Currently seconded to National ICT Australia Limited, which is funded by the Australian
  Government's Department of Communications, Information Technology and the Arts and the Australian Research Council
  through Backing Australia's Ability and the ICT Centre of Excellence Program.} \\
  {\small Department of Information Engineering, RSISE}\vspace{-2mm}\\
  {\small The Australian National University}\vspace{-2mm}\\
  {\small Canberra ACT 0200, Australia}\vspace{-2mm}\\
  {\small {\em e-mail:} Jochen.Trumpf@anu.edu.au}
}
\begin{document}
\maketitle
\begin{abstract} Superregular matrices are a class of lower triangular Toeplitz
matrices that arise in the context of constructing convolutional codes having a
maximum distance profile. These matrices are characterized by the property that
no submatrix has a zero determinant unless it is trivially zero due to the lower triangular structure.  
In this paper, we discuss how superregular matrices may be used to construct
codes having a maximum distance profile. We also introduce group actions
that preserve the superregularity property and present an upper bound on the
minimum size a finite field must have in order that a superregular matrix of a
given size can exist over that field.
\\

\end{abstract}

\section{Introduction}
Convolutional codes are a class of error-correcting codes that have enjoyed
wide use in practical applications due to the existence of efficient
non-algebraic decoding algorithms. From a mathematical standpoint, however, the
situation is still rather unsatisfying, as there are relatively few algebraic
constructions of convolutional codes with provably good distance properties or
which can be algebraically decoded.  Recent years have seen interesting
developments in the algebraic theory of convolutional codes: the papers~\cite{Heide2,Heide4,Heide5,gl02u,14}
extend  the notion of cyclicity familiar from block code theory to
convolutional codes; the papers~\cite{Heide3,Heide6} investigate
weight enumerators and the existence of a MacWilliams Identity for convolutional
codes; the paper~\cite{ro99a} uses methods from
systems theory to construct convolutional codes having a designed distance; and the papers~\cite{Heide1,gl03r,12,ro99a1,sm01a} provide results concerning
convolutional codes having certain maximal distance properties.  Motivated by existence results appearing in certain members of this last set of papers, we
decided to investigate so-called {\em superregular matrices}.  These matrices arise when
one considers the problem of constructing convolutional codes having a maximum
distance profile.

The remainder of this paper is structured as follows.  In Section 2, we give a
brief introduction to convolutional codes, explain the maximal distance
properties mentioned above, and define the superregularity property.  In Section
3, we discuss how superregular matrices may be used to construct codes having a
maximum distance profile.  In Section 4, we introduce group actions that
preserve the superregularity property; these actions made it possible to reduce
the computation time necessary for performing searches for superregular matrices.  Finally, in Section 5, we present an upper bound on
the minimum field size required for a superregular matrix of a given size to
exist.

\section{Preliminaries}
In this section, we review the theory of convolutional codes relevant to the
presented work.  Let $k$ and $n$ be positive integers with $k<n$.  Let $\F$ be a
finite field of characteristic $p$, where $p$ is a prime number.
\begin{defi}
A {\em convolutional code} $\mathcal{C}$ of {\em rate} $k/n$ is a
rank-$k$ submodule of the free module $(\F [s])^n$.
\end{defi}  
Since $\F [s]$ is a principal ideal domain, $\C$ is a free module and has a well-defined rank.  It follows that a convolutional code may be viewed 
as the column space of an $n\times k$
polynomial matrix $G(s)$, the columns of which form an $\F [s]$-basis for
$\mathcal{C}$.  As a set, we have 
$$
  \C = \{v(s)=G(s)u(s) \in (\F [s])^n \, | \, u(s)\in (\F [s])^k\}.
$$ $G(s)$ is called a {\em generator matrix} for $\mathcal{C}$, the
$u(s)$ are called {\em information vectors}, and the $v(s)$ are called
{\em code vectors} or {\em codewords}.  Two generator matrices
$G_1(s)$ and $G_2(s)$ having full column rank generate the same code
if and only if there exists a $k\times k$ unimodular matrix $U(s)$
such that $G_1(s)=G_2(s)U(s)$.

The columns of $G(s)$ may be thought of as polynomials with
coefficients in $\F ^n$; we refer to the degrees of these polynomials
as the {\em column degrees} of $G(s)$ and denote the degree of the
$i$th column by $\mu_i$.  Let $m:=\max_{1\leq i\leq k}\{ \mu _i\}$.
Thinking of $(\F [s])^{(n\times k)}$ as $\F ^{(n\times k)}[s]$, we may then expand $G(s)$ into a matrix polynomial,
$$
  G(s)=G_0  + G_1s + \cdots + G_ms^m,
$$ where $G_i$ is an $n\times k$ matrix over $\F$, the entries of
which are the coefficients of $s^i$ in $G(s)$.  Similarly, thinking of $(\F [s])^k$ as $\F ^k[s]$, we may expand 
$u(s)
\in (\F [s])^k$ of degree $l$ into a vector
polynomial:
$$
  u(s)=u_0 + u_1s + \cdots + u_ls^l.
$$
We may then represent the encoding process with the multiplication
$$
  \left[\begin{array}{c}
    v_0\\
    v_1\\
    \vdots \\
    v_{l+m}
  \end{array}\right]
  =
  \left[\begin{array}{ccccc}
    G_0 & 0 & \cdots & \cdots & 0\\
    G_1 & G_0 & \ddots &  & \vdots\\
    \vdots & G_1 & \ddots & \ddots & \vdots\\
    G_{m-1} & \vdots & \ddots & \ddots & 0 \\
    G_m & G_{m-1} &  & \ddots & G_0\\
    0 & G_m & \ddots &  & G_1\\
    \vdots & \ddots & \ddots & \ddots & \vdots\\
    \vdots &  & \ddots & \ddots & G_{m-1}\\
    0 & \cdots & \cdots & 0 &   G_m
  \end{array}\right]
  \left[\begin{array}{c}
  u_0\\
  u_1\\
  \vdots \\
  u_l
  \end{array}\right],
$$ where 0 represents the $n\times k$ matrix with all entries zero.
The large matrix in the middle is called a {\em sliding generator
matrix}.  From this representation, we see the origin of the name
convolutional code: the vector coefficients of the information vector
are convoluted with the matrix coefficients of $G(s)$ to form the
codewords.

A generator matrix $G(s)$ is called {\em basic} if the only common
divisors of the determinants of its $k\times k$ submatrices belong to
$\F\, \backslash \{0\}$.  There are several characterizations of this
property; more details may be found in~\cite{gl02u,14,15}.  We note
here two of these characterizations.  The first is that a
convolutional code generated by a basic generator matrix is a direct
summand of $\F ^n[s]$.  The second is that there
exists a basic $n\times (n-k)$ matrix over $\F [s]$, $H(s)$, such that
$\C$ is the right $\F [s]$-kernel of $H^T(s)$.  Thus, an observable
code may also be described as the set
$$
  \C = \{ v(s)\in (\F [s])^n \, | \, H^T(s)v(s) = [0\,0\cdots0]^T\in (\F [s])^k \}
$$
$H(s)$ is called a {\em parity check matrix} of $\C$.  As with the generator
matrix, one may expand H(s) as the sum
$$
  H(s) = H_0 + H_1s + \cdots + H_{m'}s^{m'},
$$ where $m'$ is the largest integer such that $H_{m'} \neq 0$.  A
convolutional code $\mathcal{C}$ is called {\em observable} if one
(and hence every) generator matrix of $\mathcal{C}$ is basic.

The {\em high-order coefficient matrix} $G_{\infty}$ of
$G(s)$ is a matrix, the $i$th column of which is the vector coefficient of $s^{\mu_i}$ in
column $i$.  
A basic generator matrix for which $G_{\infty}$ has full rank is
called {\em minimal}.  
An important invariant of a convolutional code is its {\em complexity}, defined as follows:
\begin{defi}
The {\em complexity} $\delta$ of a convolutional code $\mathcal{C}$ is
the maximum of the degrees of the (polynomial) determinants of the $k\times k$
submatrices of any generator matrix of $\mathcal{C}$.
\end{defi}
This definition makes sense, as the equivalence relation described above
preserves the degrees of these determinants.  A code of rate $k/n$ and complexity $\delta$
will also be referred to as an $(n,k,\delta)$-code.  If the code is observable, the
complexity is normally referred to as the {\em degree} of the code;
see~\cite{ro99a1} for the geometric motivation for this terminology.    

In general, $G_{\infty}$ need not have full rank; in this case,
$\sum_{i=1}^k \mu_i >\delta$.  It is shown in~\cite{17} that it is
always possible to find a unimodular matrix $U(s)$ such that
$G(s)U(s)$ has a full-rank high-order coefficient matrix and a set
$\{\nu_i\}_{i=1}^k$ of column degrees satisfying
$\nu_1\leq\nu_2\leq\cdots\leq\nu_k$ and $\sum_{i=1}^k \nu_i =\delta$.
In this case, the column degrees are invariants of the code generated
by $G(s)$ and are called the {\em column indices} or the {\em
  Kronecker indices} of the code.  If $G(s)$ is minimal, the column
degrees $\nu_1\leq\nu_2\leq\cdots\leq\nu_k$ mentioned above are called
the {\em minimal column indices} or the {\em Forney indices} of the
code.  For the remainder of the paper, we assume all codes to be
observable and generator matrices to be minimal.

The following truncated sliding generator matrices $G^c_j\in\F^{(j+1)n\times(j+1)k}$
and parity check matrices $H^c_j\in\F^{(j+1)n\times(j+1)(n-k)}$ will be of
importance in this work.  These are defined for each $j\in \N _0$ as
\begin{equation}\label{e-Gcj}
\begin{array}{rcl}
       G^c_j := &\begin{bmatrix}
                 G_0& 0& \cdots &0\\
                 G_1& G_0& \ddots &\vdots\\
                 \vdots & \vdots & \ddots &0\\
                 G_j&G_{j-1}& \cdots&G_0
              \end{bmatrix}\,\,\mbox{and}\,\,
       H^c_j := \begin{bmatrix}
                 H_0& H_1&\cdots&H_j      \\
                 0& H_0& \cdots & H_{j-1}      \\
                 \vdots& \ddots & \ddots &\vdots\\       
                 0&\cdots& 0 &H_0
              \end{bmatrix},
\end{array}
\end{equation}
where $G_j=0$ ($H_j=0$) if $j>m$ ($j>m'$).  The relation $H^T(s)G(s) =
0$ immediately implies that $(H^c_j)^TG^c_j = 0$ for all $j\in \N _0$.

We turn now to some distance notions for convolutional codes. 
\begin{defi}
Let $x \in \F ^n$.  The {\em Hamming weight} of $x$, wt($x$), is the
number of nonzero components of $x$.  Let $v(s)\in \F ^n[s]$ be given
by $v(s):=v_0 + v_1s + \cdots + v_ls^l$ for some nonnegative integer
$l$.  The {\em weight} of $v(s)$, wt($v(s)$), is the sum of the
weights of its $\F ^n$-coefficients:
$$
  \mbox{wt}(v(s)):=\sum_{i=0}^l\mbox{wt}(v_i).
$$
\end{defi}
We then have
\begin{defi}
Let $\C$ be an $(n,k,\delta)$-code.  Then, the {\em free distance} of
$\C$, $d_{free}(\C)$, is
$$
  d_{free}(\C):=\min_{v(s)\in \mathcal{C}}\{\mbox{wt}(v(s)) \, | \, v(s) \not = 0\}.
$$
\end{defi}
The following theorem gives an upper bound for how large the free
distance of an $(n,k,\delta)$-code can be:
\begin{theo}\label{Singleton}
Let $\C$ be an $(n,k,\delta)$-code.  Then, $d_{free}(\C)$ satisfies
$$
  d_{free}(\C)\leq (n-k)\Big(\Big\lfloor\frac {\delta}{k} \Big\rfloor +1\Big) +\delta +1.
$$
\end{theo}
This bound is proven in~\cite{ro99a1}.  It is known as the {\em
generalized Singleton bound}.  The reason for this is that a
convolutional code of complexity 0 is a block code, as it has a
generator matrix with all entries in $\F$.  If we set $\delta =0$ in
the above expression, it reduces to the Singleton bound from the
theory of block codes.  In analogy with the block code case,
convolutional codes having a free distance meeting the generalized
Singleton bound are called {\em maximum distance separable} ({\em
MDS}).

The free distance is a global distance measure and determines the
maximum number of errors that may be introduced to a codeword without
jeopardizing correct decoding.  A more local distance measure, relevant 
to the performance of {\em sequential decoding algorithms}
(see, for example,~\cite{Costello}), is given by column distances.
These are defined as follows:
\begin{defi}
Let $\mathcal{C}$ be a convolutional code.  For $j\in \N _0$ and a
polynomial vector $v(s)\in \F ^n[s]$ of degree $l$, set
$v_{[0,j]}(s):=v_0+v_1s+\cdots +v_js^j$ (where $v_j=0$ if $j>l$).  Then, the {\em $j$th column
distance} of $\mathcal{C}$, $d_j^c(\mathcal{C})$, is defined as
$$
d_j^c(\mathcal{C}):=\min _{v(s)\in \mathcal{C}}\{\wt(v_{[0,j]}(s)) \,
| \, v_0\neq 0 \}.
$$
Because of the assumption that $\C$ is observable, the fact that $v_0 \neq 0$
means that the minimum is taken over codewords resulting from information
vectors with $u_0 \neq 0$.
It is easy to see that the column distances satisfy 
\begin{equation} 
\label{e-dist.inequ}
  d^c_0(\mathcal{C})\leq d^c_1(\mathcal{C})\leq d^c_2(\mathcal{C})\ldots 
  \leq \lim_{j\rightarrow\infty}d^c_j(\mathcal{C})=d_{free}(\C).
\end{equation}
\end{defi}
The $(m +1)$-tuple of numbers $(d^c_0(\C), d^c_1(\C),\ldots,d^c_m(\C))$ is called the
{\em column distance profile} of the code~\cite[p. 112]{jo99}.

The following theorem gives an upper bound for the $j$th column distance:
\begin{theo}                  \label{P-dcj.bound}
Let $\mathcal{C}$ be an $(n,k,\delta)$-code.  For every $j\in \N _0$, we have
\[
d_j^c(\mathcal{C})\leq(n-k)(j+1)+1.
\]
If $d_j^c(\mathcal{C})=(n-k)(j+1)+1$ for some $j$, then $d_i^c(\mathcal{C})=(n-k)(i+1)+1$ 
when $i\leq j$.
\end{theo}
A proof may be found in~\cite{gl03r}. By considering the bound in
Theorem~\ref{Singleton}, one easily sees that the largest integer for
which the upper bound in Theorem~\ref{P-dcj.bound} can be attained is
$L:=\lfloor\frac{\delta}{k}\rfloor +
\lfloor\frac{\delta}{n-k}\rfloor$.  An $(n,k,\delta)$-code $\C$ is
said to be {\em maximum distance profile} ({\em MDP}) if $d_L^c(\C) =
(L + 1)(n-k) + 1$; note that Theorem~\ref{P-dcj.bound} implies that
$d_j^c(\C) = (j + 1)(n-k) + 1$ for all $j\in\{0,1,\ldots ,L\}$.  In
other words, the column distances of an MDP code are maximal for as
long as possible.

We close this section by introducing superregular matrices; in Section
2, we will see their relevance to the construction of MDP codes.
\begin{defi}\label{SRMDef}
Let $A$ be the $(\gamma +1) \times (\gamma +1)$ lower triangular Toeplitz
      matrix  
      $$
                \begin{bmatrix}
          a_0    & 0      & \cdots  & 0        \\
          a_1    & a_0    & \ddots  & \vdots   \\
          \vdots & \ddots & \ddots  & 0        \\
          a_{\gamma}    & \cdots & a_1     & a_0
          \end{bmatrix}.
      $$
      Let $s\in \{1,2,\ldots ,\gamma +1\}$.  Suppose that $I:=\{ i_1,\ldots ,i_s \}$ is a set of row indices of $A$,
      $J:=\{ j_1,\ldots ,j_s \}$ is a set of column indices of $A$, and that that the elements of each set are
      ordered from least to greatest.  We denote by $A^{i_1,\ldots ,i_s}_{j_1,\ldots ,j_s}$
      the submatrix of $A$ formed by intersecting the columns indexed by the members of $J$ and the rows indexed by
      the members of $I$.  A submatrix of $A$ is said to be {\em proper} if, for each $\nu\in \{ 1,2,\dots ,s \}$, the
      inequality $j_\nu\leq
      i_\nu$ holds.  The matrix $A$ is said to be {\em superregular} if every proper submatrix of $A$ has a nonzero determinant.
\end{defi}
\begin{rem}\label{SRMRem}

Observe that the proper submatrices of $A$ are the
only ones that can possibly have a nonzero determinant: If
$j_{\nu}>i_{\nu}$ for some $\nu$, then, in the submatrix
$A^{i_1,\ldots,i_s}_{j_1,\ldots,j_s}$, the upper right block
consisting of the first $\nu$ rows and the last $s-\nu+1$ columns contains only zero entries.  Hence, the matrix formed from the first $\nu$ rows
of $A^{i_1,\ldots,i_s}_{j_1,\ldots,j_s}$ can have rank at most
$\nu-1$.  For example, if $\gamma \geq 4$, we can consider the
submatrix
      \begin{equation*}
        A^{1,2,5}_{1,3,4}=\begin{bmatrix} a_{0}&0&0\\ a_{1}&0&0\\
        a_{4}&a_{2}&a_{1}\end{bmatrix}.
      \end{equation*}
      This submatrix clearly has a zero determinant regardless of what
      $a_0,a_1,a_2$, and $a_4$ are.
\end{rem}

\section{A Construction of MDP Codes from Superregular Matrices}
In this section, we take a closer look at the transposes $(H^c_j)^T \in \F ^{(j+1)(n-k)\times (j+1)n}$ of the
matrices that were introduced in
(\ref{e-Gcj}):
$$
    (H^c_j)^T = \begin{bmatrix}
                 H_0^T& 0&\cdots&0      \\
                 H_1^T& H_0^T& \ddots & \vdots      \\
                 \vdots& \vdots & \ddots &0\\       
                 H_j^T&H_{j-1}^T& \cdots &H_0^T
              \end{bmatrix}.
$$
We then have the following definition:
\begin{defi}
$(H^c_j)^T$ is said to have the 
{\em maximum span property} if none of the first $n$ columns of $(H^c_j)^T$ is
contained in the span of any other $(j+1)(n-k)-1$ columns
of $(H^c_j)^T$. 
\end{defi}
We note here that, of course, each of the first $n$ columns of $(H^c_j)^T$ is in the span of some set of 
$(j+1)(n-k)$ columns of $(H^c_j)^T$.  We also make the observation that, if $\C$ is an $(n,k,\delta)$-code and
$(H^c_j)^T$ the transpose of its $j$th truncated sliding parity check matrix, then $d^c_j(\C)=(n-k)(j+1)+1$ if and only
if $(H^c_j)^T$ has the maximum span property.  In particular, $\C$ is an MDP code if and only if $(H^c_L)^T$ has
the maximum span property. 

Since $H(s)$ is basic, we may assume without loss of generality (for the purpose of what follows) that the
$(n-k)\times (n-k)$ submatrix of $H_0$ consisting of the first $n-k$ columns has full rank.  In this case, left
multiplication by an invertible matrix followed by a column permutation gives the matrix
$$
         \widehat {(H^c_j)^T}:=\left[\!\begin{array}{cccc|cccc}
                   1& 0&\cdots& 0 &\widehat H_0    & 0      & \cdots & 0 \\
                   0& 1&\ddots      & \vdots & \widehat H_1    &\widehat H_0   & \ddots & \vdots \\
                   \vdots &\ddots  &\ddots& 0 & \vdots & \ddots &\ddots & 0\\
                   0&\cdots  & 0     &1 &\widehat H_j& \widehat H_{j-1} &\cdots&\widehat H_0\end{array}\!\right] := \left[
		   I_{(j+1)(n-k)} \, | \, \widehat H \right].
$$
Recalling Definition 3.1, we will also say that any matrix having the same form as $\widehat {(H^c_j)^T}$ has the maximum span property if none of the first $k$
columns of $\widehat H$ is contained in the span of any other $(n-k)(j+1)-1$ columns of $\widehat {(H^c_j)^T}$. 
Note that if $(H^c_j)^T$ has the maximum span property, then so does $\widehat {(H^c_j)^T}$.  The next theorem is a
simplified version of~\cite[Theorem 3.5]{gl03r}; it relates the superregularity property to the maximum span
property:
\begin{theo}\label{T-superreg}
Let $T$ be a  $(j+1)\times (j+1)$ lower triangular Toeplitz 
matrix:
\begin{equation}\label{e-HToeplitz}
    T:=
    \begin{bmatrix} t_0    & 0      & \cdots & 0 \\
                    t_1    & t_0    & \ddots & \vdots  \\
                    \vdots & \ddots &\ddots & 0\\
                    t_j    & t_{j-1} &   \cdots  & t_0\end{bmatrix}
\end{equation}  
 Then, $T$ is superregular if and only if the matrix $[I_{j+1} \, | \, T]$ has the maximum span property.
\end{theo}
Given $k$, $n$, and $j$, we will now see how to use an appropriately sized superregular matrix to construct a
matrix which has the form of $\widehat {(H^c_j)^T}$ and the maximum span property (for more details,
see~\cite[Appendix C]{gl03r}):

\begin{theo} \label{result}
Let $T$ be an $l\times l$ superregular matrix, where 
$l\geq (j+1)(n-1)$.  
Let $T'$ the submatrix obtained from $T$ by intersecting the rows indexed by $\{k,
k+1,\ldots, n-1, n-1+k, n-1+k+1, \ldots, 2(n-1),
\ldots ,j(n-1)+k,j(n-1)+k+1,\ldots,j(n-1)+n-1\}$ and columns indexed by $\{1,2,\ldots, k, n-1+1,n-1+2,\ldots, n-1+k,
\ldots,j(n-1)+1,j(n-1)+2,\ldots,j(n-1)+k\}$. Then, $T'\in \F ^{(j+1)(n-k)\times (j+1)k}$ is a lower
block triangular Toeplitz matrix such that $[I_{(j+1)(n-k)} \, | \, T']$ has the maximum span property. 
\end{theo}
If we let $j=L$, we may use the resulting matrix to construct a parity check matrix of an MDP convolutional code.

\begin{exa}
In this example, we consider the following matrix over $\F _{64}$:
$$
     \begin{bmatrix}1&0&0&0&0&0&0&0\\ \omega&1&0&0&0&0&0&0\\
                   \omega^9&\omega&1&0&0&0&0&0\\\omega^{33}&\omega^9&\omega&1&0&0&0&0\\
                   \omega^{33}&\omega^{33}&\omega^9&\omega&1&0&0&0\\
                   \omega^9&\omega^{33}&\omega^{33}&\omega^9&\omega&1&0&0\\
                   \omega&\omega^9&\omega^{33}&\omega^{33}&\omega^9&\omega&1&0\\
                   1&\omega&\omega^9&\omega^{33}&\omega^{33}&\omega^9&\omega&1
     \end{bmatrix}.
$$
Here, $\omega$ is a root of $x^6+x+1 \in \F _2[x]$ and is thus a primitive field element.  According to~\cite{gl03r},
this matrix is superregular.  We let $k=2$, $n=3$, and $\delta =2$.  Then, $L=3$, and, since $(L+1)(n-1)=8$, we
may use Theorem~\ref{result} to form the matrix
$$
  \left[
    \begin{array}{cccc|cccccccc}
    
      1 & 0 & 0 & 0 & \omega & 1 & 0 & 0 & 0 & 0 & 0 & 0\\
      0 & 1 & 0 & 0 & \omega ^{33} & \omega ^{9} & \omega & 1 & 0 & 0 & 0 & 0\\
      0 & 0 & 1 & 0 & \omega ^9 & \omega ^{33} & \omega ^{33} & \omega ^9 & \omega & 1 & 0 & 0\\
      0 & 0 & 0 & 1 & 1 & \omega & \omega ^9 & \omega ^{33} & \omega ^{33} & \omega ^9 & \omega & 1
    \end{array}
  \right],
$$
which has the maximum span property.  We may then use this matrix to construct a polynomial generator matrix for
an MDP $(3,2,3)$ code.  We could, for example, think of the $1\times 2$ matrices making up the first block column
of the right-hand side of this matrix as a finite sequence of Markov parameters and compute a minimal partial
realization of this sequence (see, for example,~\cite{an86}).  We would then compute the corresponding transfer
function.  After some small algebraic manipulations and fixing an ordering for the codeword components, we would
arrive at a polynomial generator matrix and its corresponding
convolutional code (see~\cite{ro99a} for more information about the relationship between the linear systems and
polynomial representations of convolutional codes).  In this example, one possibility for a resulting generator matrix is
$$
  \begin{bmatrix}
    s^2 + \omega ^{57}s + \omega ^{62} & 0\\
    0                                  & s^2 + \omega ^{57}s + \omega ^{62}\\
    \omega s^2+\omega ^{44} s+\omega ^{54}       & s^2 +\omega ^{17} s+\omega ^{21}
  \end{bmatrix}.
$$ 
\end{exa}
\section{Group Actions Preserving the Superregularity \\Property}
In this section, we consider group actions that allow one to create new
superregular matrices from a given one.  The main utility of these actions is
that they reduce the number of matrices one must check when performing a computer
search for matrices having the superregularity property.  

For a prime power $p^e$ and a nonnegative integer
$\gamma$, denote by $SR(p^e,\gamma )$ the set of superregular matrices of
dimension $\gamma +1$ over $\F _{p^e}$.  The first group action is a corollary of the following result,
which is proven in~\cite{gl03r}:
\begin{theo}\label{Z2}
Suppose that $A\in SR(p^e,\gamma )$.  Then, $A^{-1}\in SR(p^e,\gamma )$.
\end{theo}
\begin{cor}\label{Z2Cor}
There is an action $*_1$ of the additive group $(\Z ,+)$ on the set $SR(p^e,\gamma )$, given by
$$
  \begin{array}{cccc}\label{Action1}
    *_1: & \Z \times SR(p^e,\gamma ) & \longrightarrow & SR(p^e,\gamma )\\
         & \hspace{-0.7cm}(x,A)                         & \longmapsto     &
         A^{(-1)^{x}}.
  \end{array}
$$
\end{cor}
\begin{proof}
By Theorem~\ref{Z2}, $A^{-1}$ is superregular, and it is clear that $*_1$ is a
group action.
\end{proof}
We next have the following simple result:
\begin{lemma}\label{SRInverse}
If $A\in SR(p^e,\gamma )$ and $\gamma \geq 2$, then $A\neq A^{-1}$.
\end{lemma}
\begin{proof}
     Suppose
     $$
          A:=
          \begin{bmatrix}
          a_0    & 0      & \cdots  & 0        \\
          a_1    & a_0    & \ddots  & \vdots   \\
          \vdots & \ddots & \ddots  & 0        \\
          a_{\gamma}    & \cdots & a_1     & a_0
          \end{bmatrix}
     $$
     is a superregular matrix, where $\gamma \geq 2$.  Suppose in addition that $A=A^{-1}$.  Then, since
     $A^2=I_{\gamma +1}$, we must have that $a_0a_1+a_1a_0=2a_0a_1=0$.  By
     hypothesis, $a_0, a_1\not =0$.  Thus, the field must have
     characteristic 2.  We must also have that
     $a_2a_0+a_1^2+a_0a_2=2a_0a_2+a_1^2=0$.  Since the field has
     characteristic 2, this equation reduces to $a_1^2=0$.  This is a
     contradiction, as $a_1\not =0$.  Thus, $A\not =A^{-1}$.  
\end{proof}
\begin{cor}\label{SREven}
If $\gamma \geq 2$, then $|SR(p^e,\gamma )|$ is even. 
\end{cor}

We next describe an action of the multiplicative group $\F _{p^e}^*:=\F _{p^e} \backslash \{
0 \}$ of nonzero field elements of $\F _{p^e}$ on $SR(p^e,\gamma )$:
\begin{theo}\label{Fp*}
     Let $\alpha\in\F _{p^e}^*$.  Define  
     $$
          \alpha \bullet A:=
          \begin{bmatrix}
                    a_0    & 0          & \cdots        & \cdots     &
          0      \\
          \alpha    a_1    & \ddots        & \ddots        &            &
          \vdots \\
          \alpha ^2 a_2    & \ddots & \ddots           & \ddots     &
          \vdots \\
                    \vdots & \ddots     & \ddots        & \ddots     &
          0      \\
          \alpha ^{\gamma} a_{\gamma}    & \cdots     & \alpha ^2 a_2 & \alpha a_1 &
          a_0
          \end{bmatrix}
     $$
     Then, the map
     $$
       \begin{array}{cccc}\label{Action2}
         *_2: & \F _{p^e}^* \times SR(p^e,\gamma ) & \longrightarrow & SR(p^e,\gamma )\\
              & \hspace{-0.8cm}(\alpha,A)                         & \longmapsto     &
              \alpha \bullet A
       \end{array}
     $$
     is an action of $\F _{p^e}^*$ on $SR(p^e,\gamma )$.
\end{theo}
\begin{proof}
It is readily verified that $*_2$ is a group action.  Let 
$$
     D:=
     \begin{bmatrix}
       1       &  0       &  \cdots  &  0          \\
       0       &  \alpha  &  \ddots  &  \vdots     \\
       \vdots  &  \ddots  &  \ddots  &  0          \\
       0       &  \cdots  &  0       &  \alpha ^{\gamma}
     \end{bmatrix}.
$$
One can then describe $*_2$ through conjugation by $D$:  $\alpha \bullet A = DAD^{-1}$.  This changes the determinants of
submatrices of $A$ only by factors of powers of $\alpha$, and it follows that
$\alpha \bullet A$ is also superregular.
\end{proof}

The final group action we consider makes use of the Galois group
$Aut_{\F_p}\F_{p^e}$.  We recall here the fact that $Aut_{\F_p}\F_{p^e} = \langle \sigma \rangle$, where $\sigma
\in Aut_{\F _p}\F _{p^e}$ is the $\F _p$-automorphism defined by $\sigma (x) = x^p\,\, \forall x\in \F _{p^e}$.  
Consequently, $Aut_{\F _p}\F _{p^e}\cong
\Z/e\Z$.
\begin{theo}\label{ZkZ}
     Let $A\in SR(p^e,\gamma )$. Let $i\in \Z /e\Z$.  Define
     $$
          i \circ A:=
          \begin{bmatrix}
                    a_0^{p^i}    & 0          & \cdots        & \cdots     &
          0      \\
              a_1^{p^i}    & \ddots        & \ddots        &            &
          \vdots \\
           a_2^{p^i}    & \ddots & \ddots           & \ddots     &
          \vdots \\
                    \vdots & \ddots     & \ddots        & \ddots     &
          0      \\
           a_{\gamma}^{p^i}    & \cdots     & 
          a_2^{p^i} &  a_1^{p^i} &
          a_0^{p^i}
          \end{bmatrix}.
     $$
     Then, the map
     $$
       \begin{array}{cccc}\label{Action3}
         *_3: & \Z /e\Z \times SR(p^e,\gamma ) & \longrightarrow & SR(p^e,\gamma )\\
              & \hspace{-0.8cm}(i,A)                         & \longmapsto     &
              i \circ A
       \end{array}
     $$
     is an action of the additive group $\Z /e\Z$ on $SR(p^e,\gamma )$.
\end{theo}
\begin{proof}
     It is readily verified that $*_3$ is a group action.  As the entries of $A$ belong to a field of characteristic $p$, the
     determinant of a submatrix of $i \circ A$ is the determinant of the
     corresponding submatrix of $A$ raised to the $p^i$th power.  Thus, $i \circ A$
     is also superregular.
\end{proof}

\section{An Upper Bound for the Required Field Size}
An important question to consider in trying to better understand superregular
matrices is that of how large a finite field must
be in order that a superregular matrix of a given size can exist over that
field.  For example, no $3\times 3$ superregular matrix exists over the field
$\F _2$:  all entries in the lower triangular part of a superregular matrix must be nonzero, which means that
in this case all such entries would have to be 1; clearly, this does not result in a
superregular matrix, since the lower left $2\times 2$ submatrix has a zero determinant.  In this section, we give an upper bound on the required
field size.  We begin with a technical lemma:
\begin{lemma}\label{Bijection}
     Let $i$ be a nonnegative integer and $\gamma$ a positive integer.  
     Let $S_{i+1}$ denote the set of integer sequences $\{s_l\}_{l=0}^{i+1}$,
     where $s_0=0$, $s_{i+1}=\gamma$, and $s_l<s_{l+1}\, \forall \, l\in
     \{0,1,\ldots ,i\}$.
     Let $S_{i,\gamma }:=\{\{s_l\}_{l=0}^{i+1}\in S_{i+1} \,\,  with \,\,   
     s_j+s_{i-j+1}\leq \gamma ,\, j=0,1,\ldots ,\lceil\frac{i}{2}\rceil\}$.  
     Let $T_{i+1}$ denote the set of integer sequences $\{t_l\}_{l=0}^{i+1}$,
     where $t_0=0$, $t_{i+1}=\gamma$, and $t_l<t_{l+1}\, \forall \, l\in
     \{0,1,\ldots ,i\}$.  Let
     $T_{i,\gamma }:=\{\{t_l\}_{l=0}^{i+1}\in T_{i+1} \,\, with \,\, \sum_{l=0}^{m}
     (-1)^l (t_{l+1}-t_l)\geq 0, \, m=0,1,\ldots ,i\}$.
     Then, 
     $S_{i,\gamma }$ and $T_{i,\gamma }$ are finite sets, and $|S_{i,\gamma
     }|$=$|T_{i,\gamma }|$.    
\end{lemma}
\begin{proof}
     It is clear that $S_{i,\gamma }$ and $T_{i,\gamma }$ are finite
     sets.  We now proceed to prove the second part of the claim.  We will do this by first constructing an
     injective map from $S_{i,\gamma }$ to $T_{i,\gamma }$ and then an injective map from $T_{i,\gamma }$ to
     $S_{i,\gamma }$.  
     Throughout the proof, $t_{-1}$ and $s_{-1}$ are defined to be 0.
     Suppose first that $i$ is even.  Let $\{s_l\}_{l=0}^{i+1} \in
     S_{i,\gamma }$.  For $j\in \{0,1,\ldots ,\frac{i}{2} \}$, we make
     the recursive definitions
     \begin{align*}
     t_{2j+1}&:=t_{2j}+s_{i-j+1}-s_{i-j},\\ 
     t_{2j}&:=t_{2j-1}+s_{j}-s_{j-1}.  
     \end{align*}
     Note that $t_0=s_0=0$.  
     Then, for $m\in \{0,1,\ldots
     ,i\}$, we have    
     \begin{align*}
     \sum_{l=0}^{m}
          (-1)^l (t_{l+1}-t_l) &=t_1-(t_2-t_1)+(t_3-t_2)-\cdots
          +(-1)^{m}(t_{m+1}-t_m)\\
          &=2t_1-2t_2+2t_3-\cdots +(-1)^{m}t_{m+1}\\
          &=2t_1-2t_2+2t_3-\cdots
          -t_{2j}\\
          &\hspace{3cm} \mbox{or}\\
          &=2t_1-2t_2+2t_3-\cdots +t_{2j+1}, 
     \end{align*}
     where $j\in \{0,1,\ldots
     ,\frac{i}{2} \}$ is $(m-1)/2$ or $m/2$ as $m$ is odd or even, respectively.  Suppose
     first that $m$ is odd.  The sum then becomes
     \begin{align*}
          2t_1-2t_2+2t_3-\cdots
          -t_{2j}&=2t_1-2(t_1+s_1)+2t_3-\cdots
          -(t_{2j-1}+s_j-s_{j-1})\\
          &=t_{2j-1}-(s_j-s_{j-1})-2s_{j-1}\\
          &=t_{2j-1}-s_j-s_{j-1}.
     \end{align*}
     Using the definition of $t_{2j+1}$ with $j$ replaced by $j-1$, we may write
     $$
          t_{2j-1}-s_j-s_{j-1}=t_{2j-2}+s_{i-j+2}-s_{i-j+1}
          -s_j-s_{j-1}.
     $$    
     Using again the definitions above, we see that, for
     any integer $h\in \{0,\ldots ,j-1 \}$, 
     $$
          t_{2j-2}+s_{i-j+2}-s_{i-j+1}-s_j-s_{j-1}=
          t_{2(j-h)-2}+s_{i-(j-(h+1))+1}-s_{i-j+1}-s_j-s_{j-(h+1)}:
     $$
     this equation holds trivially if $h=0$, and, if $h\in \{ 0,\ldots ,j-2 \}$, we have
     \begin{align*}
          & t_{2(j-h)-2}+s_{i-(j-(h+1))+1}-s_{i-j+1}-s_j-s_{j-(h+1)}=\\
          &
t_{2(j-h-1)-1}+s_{j-h-1}+s_{(j-h-1)-1}+s_{i-(j-(h+1))+1}-s_{i-j+1}-s_j-s_{j-(h+1)}
          =\\&
t_{2(j-h-2)}+s_{i-(j-h-2)+1}-s_{i-(j-h-2)}+s_{j-h-1}-s_{(j-h-1)-1}+s_{i-(j-(h+1))+1}
          \\& -s_{i-j+1}-s_j-s_{j-(h+1)}=\\& t_{2(j-(h+1))-2}+s_{i-(j-((h+1)+1))+1}
          -s_{i-j+1}-s_j-s_{j-((h+1)+1)}.
     \end{align*}
     In particular, letting $h=j-2$ and recalling that $s_{i+1}=\gamma$ and $s_0=0$,
     we see that
     $$
          t_{2j-2}+s_{i-j+2}-s_{i-j+1}-s_j-s_{j-1}=\gamma -s_{i-j+1}-s_j.
     $$
     Because $\{s_l\}_{l=0}^{i+1} \in S_{i,\gamma }$, $\gamma -s_{i-j+1}-s_j\geq 0$.  
     
     Suppose now that $m$ is even.  We are then interested in the sum
     $$
          2t_1-2t_2+2t_3-\cdots +t_{2j+1}.
     $$
     From the analysis in the case of $m$ odd, we see that we may write
     \begin{align*}
          2t_1-2t_2+2t_3-\cdots
          +t_{2j+1}&=t_{2j-1}-s_j-s_{j-1}-t_{2j}+t_{2j+1}\\
          &=t_{2j-1}-s_j-s_{j-1}-t_{2j}+t_{2j}+s_{i-j+1}-s_{i-j}\\
          &=t_{2j-1}-s_j-s_{j-1}+s_{i-j+1}-s_{i-j}.
     \end{align*}
     Using the same method as before, we see that, for any integer
     $h\in \{0,\ldots ,j-1 \}$,
     $$
          t_{2j-1}-s_j-s_{j-1}+s_{i-j+1}-s_{i-j}=t_{2(j-h)-1}
          +s_{i-(j-h)+1}-s_{i-j}-s_j-s_{j-(h+1)}.
     $$
     In particular, letting $h=j-1$, the right side of this equality reduces to
     $$
          t_1+s_i-s_{i-j}-s_j.
     $$
     Using the definition of $t_1$, this becomes
     \begin{align*}
          s_{i+1}-s_i+s_i-s_{i-j}-s_j&=s_{i+1}-s_{i-j}-s_j\\
          &=\gamma -s_{i-j}-s_j.
     \end{align*}
     Since $s_{i-j}<s_{i-j+1}$ and $\gamma -s_{i-j+1}-s_j\geq 0$, it follows that $\gamma -s_{i-j}-s_j\geq 0$.
     
     It
     remains to be shown that $t_{i+1}=\gamma$.  We have
     \begin{align*}
          t_{2j+1}&=t_{2j}+s_{i-j+1}-s_{i-j}\\
          &=t_{2j-1}+s_j-s_{j-1}+s_{i-j+1}-s_{i-j}\\
          &=t_{2j-2}+s_{i-j+2}+s_j-s_{j-1}-s_{i-j}.
     \end{align*}
     In the same way as above, one sees that, for any integer $h\in
     \{1,\ldots ,j \}$, 
     $$
          t_{2j-2}+s_{i-j+2}+s_j-s_{j-1}-s_{i-j}=t_{2(j-h)}+s_{i-(j-h)+1}
          -s_{j-h}+s_j-s_{i-j}.
     $$
     In particular, letting $j=\frac{i}{2}$ and $h=j$, the right side of the
     preceding equality simplifies to $s_{i+1}$.  By hypothesis, $s_{i+1}=\gamma$. 
     Thus, $t_{i+1}=\gamma$.
     This completes the case of $i$ even.  
     
     Suppose now that $i$ is odd.  For $j\in \{0,1,\ldots
     ,\frac{i-1}{2} \}$, define $t_{2j+1}$ as above.  For $j\in
     \{0,1,\ldots ,\frac{i+1}{2} \}$, define $t_{2j}$ as above.  The
     proof then proceeds as in the case of $i$ even.
     
     We thus obtain a well-defined map $f$ from $S_{i,\gamma }$ to
     $T_{i,\gamma }$: if $\{s_l\}_{l=0}^{i+1} \in S_{i,\gamma }$, let
     $f(\{s_l\}_{l=0}^{i+1})=\{t_l\}_{l=0}^{i+1}$, where each $t_l$ is
     defined via the equations at the beginning of the proof.  It
     follows immediately from the definition of the $t_l$ that $f$ is
     injective.  Thus, to show $|S_{i,\gamma }|$=$|T_{i,\gamma }|$, it
     will suffice to construct a similar injective map from
     $T_{i,\gamma }$ to $S_{i,\gamma }$.
     
     Suppose first that
     $i$ is even.  Let $\{t_l\}_{l=0}^{i+1} \in T_{i,\gamma }$.  For $j\in \{0,1,\ldots
     ,\frac{i}{2} \}$, we make the recursive definitions
     \begin{align*}
     s_j&:=s_{j-1}
     +t_{2j}-t_{2j-1}\\
     s_{i-j}&:=s_{i-j+1}-t_{2j+1}+t_{2j}.
     \end{align*}
     Note that $s_0=t_0=0$.  
     Then, using these definitions, we have that
     \begin{align*}
          s_j+s_{i-j+1}&=s_{j-1}+t_{2j}-t_{2j-1}+s_{i-j+2}+t_{2j-2}-t_{2j-1}
          \\
          &= s_{j-2}+t_{2j-2}-t_{2j-3}+t_{2j}-t_{2j-1}+s_{i-j+3}+t_{2j-4}
          -t_{2j-3}+t_{2j-2}-t_{2j-1}\\
          &=\\
          &\hspace{0.25cm}\vdots\\
          &=\gamma -2t_1+2t_2-\cdots +t_{2j}.
     \end{align*}
     Because $\{t_l\}_{l=0}^{i+1} \in T_{i,\gamma }$,
     $2t_1-2t_2+\ldots -t_{2j}\geq 0$.  Thus, $s_j+s_{i-j+1}\leq
     \gamma$.  It remains to be shown that $s_{i+1}=\gamma$.  From the
     definition of $s_0$, it is clear that $s_0=0$.  We have that
     \begin{align*}
           s_{i+1}  & =        s_i+t_1\\
                    & =        s_{i-1}+t_3-t_2+t_1\\
                    & =        s_{i-2}+t_5-t_4+t_3-t_2+t_1\\
                    & \,\,\,\, \vdots   \\
                    & =        s_{\frac{i}{2}}+t_{i+1}-t_i+t_{i-1}-t_{i-2}+
                               \cdots +t_3-t_2+t_1.  
     \end{align*}
     We also have that
     \begin{align*}
           s_{\frac{i}{2}} & = s_{\frac{i-2}{2}}+t_i-t_{i-1}\\
                           & = s_{\frac{i-4}{2}}+t_{i-2}-t_{i-3}+t_i-t_{i-1}\\
                           & \,\,\,\, \vdots   \\
                           & = -t_1+t_2-t_3+\cdots +t_{i-2}-t_{i-3}+t_i-t_{i-1}.  
     \end{align*}
     Thus, $s_{i+1}=t_{i+1}$.  Since $t_{i+1}=\gamma$ by hypothesis, it follows that
     $s_{i+1}=\gamma$.
      
     Suppose now that $i$ is odd.  For $j\in \{0,1,\ldots ,\frac{i+1}{2} \}$, 
     define $s_j$ as above.  For $j\in \{0,1,\ldots ,\frac{i-1}{2}
     \}$, define $s_{i-j+1}$ as above.  The proof then proceeds as
     in the case of $i$ even.
     
     We thus obtain a well-defined map $g$ from $T_{i,\gamma }$ to $S_{i,\gamma }$:  if $\{t_l\}
     _{l=0}^{i+1} \in T_{i,\gamma }$, let $g(\{t_l\}
     _{l=0}^{i+1})=\{s_l\}_{l=0}^{i+1}$, where each $s_l$ is defined
     via the equations above.  It follows immediately from the definition of the $s_l$ that $g$ is injective.  Thus,
     $|S_{i,\gamma }|=|T_{i,\gamma }|$.  
\end{proof}

We need one more technical lemma before computing the upper bound:
\begin{lemma}\label{Count}
     Let $\gamma$ be a positive integer.  Then, $\prod_{1\leq i\leq j\leq \gamma
     }
     \frac{2+i+j}{i+j}=\frac{1}{\gamma +2} {2(\gamma +1)\choose \gamma +1}$.
\end{lemma}
\begin{proof}
     The proof is by induction.  The claim is clearly true if $\gamma =1$.
     Suppose that $\prod_{1\leq i\leq j\leq \gamma }
     \frac{2+i+j}{i+j}=\frac{1}{\gamma +2} {2(\gamma +1)\choose \gamma +1}$ for some
     $\gamma\geq 1$.  Then,
     \begin{align*}
          &\prod_{1\leq i\leq j\leq \gamma +1}
          \frac{2+i+j}{i+j}= \prod_{1\leq i\leq j\leq \gamma }
          \frac{2+i+j}{i+j} \prod_{1\leq i\leq \gamma +1}
          \frac{i+\gamma +3}{i+\gamma +1}=\\ &\frac{1}{\gamma +2} \frac{(2\gamma
          +2)(2\gamma +1)\cdots
          1}{(\gamma +1)(\gamma )\cdots 1(\gamma +1)(\gamma )\cdots 1}
          \frac{(\gamma +4)(\gamma +5)\cdots
          (2\gamma +4)}{(\gamma +2)(\gamma +3)\cdots (2\gamma +2)}=\\
          &\frac{(2\gamma +2)(2\gamma +1)\cdots
          (\gamma +3)}{(\gamma +1)(\gamma )\cdots 1} \frac{(\gamma +4)(\gamma +5)\cdots
          (2\gamma +4)}{(\gamma +2)(\gamma +3)\cdots (2\gamma +2)}=\\
          &\frac{1}{\gamma +3}
          \frac{(2\gamma +4)(2\gamma +3)\cdots (\gamma +3)}{(\gamma +2)(\gamma +1)\cdots 1}=
          \frac{1}{\gamma +3} \frac{(2\gamma +4)(2\gamma +3)\cdots 1}{(\gamma
          +2)(\gamma +1)\cdots
          1(\gamma +2)(\gamma +1)\cdots 1}=\\ &\frac{1}{\gamma +3} {2\gamma
          +4\choose \gamma +2}=
          \frac{1}{(\gamma +1)+2} {2((\gamma +1)+1)\choose (\gamma +1)+1}.
     \end{align*} 
     This proves the claim.
\end{proof}

We are now ready to prove the main result of this section:
\begin{theo}\label{Bound}
     Let $\gamma\in \N$.  Let $C_{\gamma}$ denote the $\gamma$th Catalan number: 
     $C_{\gamma}:=\frac{1}{\gamma +1}
     {2\gamma\choose \gamma }$.  Let $\F$ be a finite field such
     that $|\F|>\frac{1}{2}\big (C_{\gamma -1} +{\gamma -1\choose \lfloor
     \frac{\gamma -1}{2} \rfloor} \big )$.  Then, there exists a $\gamma\times
     \gamma$
     superregular matrix over $\F$.
\end{theo}
\begin{proof}
     We first upper bound the number of submatrices of a
     $\gamma\times \gamma$ lower triangular Toeplitz matrix that possess a certain property and show that this 
     upper bound is given by the expression in the
     statement of the theorem.  We then observe that the number thus
     obtained is actually an upper bound on the minimal size a finite field $\F$ must
     have in order that a $\gamma\times \gamma$ superregular matrix over $\F$ can exist.

     For convenience, we drop the matrix entries with index 0.  Thus, a
     $\gamma\times \gamma$ lower triangular Toeplitz matrix $X$ with indeterminate
     entries is now defined by a first column of the form 
     $$
          [x_1\,\,\,\,\,\, x_2\,\,\,\,\,\, \cdots \,\,\,\,\,\, x_{\gamma}]^T.
     $$
     The determinants
     of the proper square submatrices of such a matrix are given by nonzero polynomials in
     these indeterminates.  Notice that $x_{\gamma}$ can appear to at most the first
     power
     in any of these polynomials; in other words, each of these polynomials has
     the property that either it is linear in $x_{\gamma}$, or $x_{\gamma}$ does not appear in
     any of its terms.  We are interested in those proper
     square submatrices whose determinants are linear in $x_{\gamma}$, and we denote the
     set of such submatrices by $L_{\gamma}$.  

     First, we observe that $L_{\gamma}$ consists of the single entry
     $x_{\gamma}=X_{1}^{\gamma}$ and 
     the submatrices $X_{1,j_1,\ldots
     ,j_{s-1}}^{i_1,i_2,\ldots ,i_{s-1},\gamma }$, where $s\in \{ 2,\ldots ,\gamma -1
     \}$, where
     $j_{\nu}\leq
     i_{\nu}$ for all $\nu\in \{1,2,\ldots ,s-1\}$.  That $X_{1}^{\gamma}$ is the only $1\times
     1$ submatrix in $L_{\gamma}$ is obvious; to see the second part, evaluate
     det$X_{1,j_1,\ldots ,j_{s-1}}^{i_1,i_2,\ldots ,i_{s-1},\gamma }$ by doing a
     cofactor expansion along the first column of $X_{1,j_1,\ldots
     ,j_{s-1}}^{i_1,i_2,\ldots ,i_{s-1},\gamma }$.  The indeterminate
     $x_{\gamma}$ appears if and only if det$X_{j_1,j_2,\ldots
     ,j_{s-1}}^{i_1,i_2,\ldots ,i_{s-1}}\not =0$.  This is the
     case if and only if $X_{j_1,j_2,\ldots
     ,j_{s-1}}^{i_1,i_2,\ldots ,i_{s-1}}$ is a proper submatrix, and
     $X_{j_1,j_2,\ldots
     ,j_{s-1}}^{i_1,i_2,\ldots ,i_{s-1}}$ is a proper submatrix if and
     only if $j_{\nu}\leq
     i_{\nu}$ for all $\nu\in \{1,2,\ldots ,s-1\}$.

     Next, we observe that $L_{\gamma }$ is closed with respect to
     transpose about the antidiagonal of $X$.  In order to see this,
     note that the transpose $\bar X_{1,j_1,\ldots
     ,j_{s-1}}^{i_1,i_2,\ldots ,i_{s-1},\gamma }$ of any submatrix
     $X_{1,j_1,\ldots ,j_{s-1}}^{i_1,i_2,\ldots ,i_{s-1},\gamma }$
     about the antidiagonal is given by $\bar X_{1,j_1,\ldots
     ,j_{s-1}}^{i_1,i_2,\ldots ,i_{s-1},\gamma }=X_{1,\gamma
     -i_{s-1}+1,\ldots ,\gamma - i_1+1}^{\gamma -j_{s-1}+1,\gamma
     -j_{s-2}+1,\ldots ,\gamma }$.  Clearly, $\gamma -i_{\nu}+1\leq
     \gamma -j_{\nu}+1 \iff j_{\nu}\leq i_{\nu}$.  Of course,
     $X_{1,j_1,\ldots ,j_{s-1}}^{i_1,i_2,\ldots ,i_{s-1},\gamma }\in
     L_{\gamma}$ is symmetric with respect to the antidiagonal of $X$
     if and only if $X_{1,j_1,\ldots ,j_{s-1}}^{i_1,i_2,\ldots
     ,i_{s-1},\gamma }=\bar X_{1,j_1,\ldots ,j_{s-1}}^{i_1,i_2\ldots
     ,i_{s-1},\gamma }$.  Let $L'_{\gamma}\subset L_{\gamma}$ denote
     those elements of $L_{\gamma}$ that are symmetric with respect to
     the antidiagonal.

     Finally, because $X$ is symmetric with respect to the antidiagonal, taking 
     the transpose $\bar X_{1,j_1,\ldots
     ,j_{s-1}}^{i_1,i_2,\ldots ,i_{s-1},\gamma }$ of a square submatrix $X_{1,j_1,\ldots
     ,j_{s+1}}^{i_1,i_2,\ldots ,i_{s-1},\gamma }$ about the antidiagonal of $X$ amounts to
     taking the transpose of $X_{1,j_1,\ldots
     ,j_{s-1}}^{i_1,i_2\ldots ,i_{s-1},\gamma }$ about its own antidiagonal.  Since the
     determinant of a matrix is the same as the determinant of its transpose about
     its antidiagonal, we have that
     det$X_{1,j_1,\ldots
     ,j_{s-1}}^{i_1,i_2,\ldots ,i_{s-1},\gamma }$= det$\bar X_{1,j_1,\ldots
     ,j_{s-1}}^{i_1,i_2,\ldots ,i_{s-1},\gamma }$.  

     By definition, the determinants of the elements of $L_{\gamma}$ are polynomials 
     linear in $x_{\gamma}$.  We wish to determine how large a finite field must
     be in order to guarantee that nonzero field
     elements may be substituted for $x_1,x_2,\ldots ,x_{\gamma}$ in such a
     way that none of these determinants is zero.  We do this by
     computing an upper bound $N_{\gamma}$ on the number of distinct polynomials giving the
     determinants of the elements of $L_{\gamma}$.  As long as the order of
     the field is bigger than $N_{\gamma}$, it is clearly possible to make all
     of these determinants nonzero.  From the above
     observations, we see that, once we have computed $|L_{\gamma}|$ and
     $|L'_{\gamma}|$, we
     may take as an upper bound 
     $N_{\gamma}:=\frac{1}{2}(|L_{\gamma}|+|L'_{\gamma}|)$.

     As seen above, the $s\times s$ submatrices in $L_{\gamma}$, $s\in
     \{ 2,\ldots ,\gamma -1 \}$, are described by sets
     $\{i_1,i_2,\ldots ,i_{s-1},j_1,j_2,\ldots ,j_{s-1}\}$ of indices,
     where $j_{\nu}\leq i_{\nu}$ for all $\nu\in \{1,2,\ldots ,s-1\}$,
     $1<i_1<\ldots <i_{s-1}<\gamma$, and $1<j_1<\ldots
     <j_{s-1}<\gamma$ (and the $1\times 1$ submatrix $x_{\gamma}$ is
     associated with an empty set of indices).  Each nonempty index
     set defines a generalized Young tableau with columns having
     height 2 and integer entries in $\{ 2,\ldots ,\gamma -1 \}$;
     conversely the entries of such a tableau constitute a set of
     indices corresponding to a submatrix in $L_{\gamma}$.  The empty
     set of indices corresponds to a tableau with all columns having
     height 0.  In~\cite{CatVien}, it is shown that the number of such
     tableaux is given by $\prod_{1\leq i\leq j\leq \gamma -2}
     \frac{2+i+j}{i+j}$.  By Lemma~\ref{Count}, this product is
     $C_{\gamma -1}$.  Consequently, $|L_{\gamma}|=C_{\gamma -1}$.

     We compute $|L'_{\gamma}|$ as follows.  If the submatrix
     $X_{1,j_1,\ldots ,j_{\gamma}}^{i_1,i_2\ldots ,i_{s-1},\gamma }\in
     L'_{\gamma}$, where $s\in \{ 2,\ldots ,\gamma -1 \}$, then, since
     $X_{1,j_1,\ldots ,j_{s-1}}^{i_1,i_2,\ldots ,i_{s-1},\gamma }=\bar
     X_{1,j_1,\ldots ,j_{s-1}}^{i_1,i_2\ldots ,i_{s-1},\gamma }$, it
     must be that $\gamma -i_{\nu}=j_{s-\nu }-1$ for all $\nu\in
     \{1,2,\ldots ,s-1\}$.  Thus, this submatrix is completely
     determined by the $s-1$ integers $w_1=j_1-1, w_2=j_2-1,\ldots
     ,w_{s-1}=j_{s-1}-1$.  Since $X_{1,j_1,\ldots
     ,j_{s-1}}^{i_1,i_2,\ldots ,i_{s-1},\gamma }\in L_{\gamma}$, we
     have $j_l=w_l+1\leq \gamma -w_{s-l}=i_l$, $l\in \{ 1,\ldots
     ,\lceil\frac{s-1}{2}\rceil \}$.  These inequalities can be
     rewritten as $w_l+w_{s-1-l}\leq \gamma -1$, $l\in \{ 1,\ldots
     ,\lceil\frac{s-1}{2}\rceil \}$.  In other words, $\{ 0,w_1,\ldots
     ,w_{s-1},\gamma -1 \}$ is a sequence that belongs to
     $S_{s-1,\gamma -1}$.  Thus, to each $s\times s$ submatrix in
     $L'_{\gamma}$, where $s\in \{ 2,\ldots ,\gamma -1 \}$, we have
     associated a unique sequence in $S_{s-1,\gamma -1}$.  Similarly,
     to each sequence in $S_{s-1,\gamma -1}$, we can associate a
     unique $s\times s$ submatrix in $L'_{\gamma}$.  If $s=1$, there
     is only the submatrix $X_{\gamma}^1$ to consider, and we have
     already associated this submatrix with the empty sequence, which
     in turn corresponds with the sequence $\{ 0,\gamma -1 \}$, the
     sole member of $S_{0,\gamma -1}$.  Thus,
     $|L'_{\gamma}|=\sum_{y=0}^{\gamma -2} |S_{y,\gamma -1}|$.  From
     Lemma~\ref{Bijection}, we know that $\sum_{y=0}^{\gamma -2}
     |S_{y,\gamma -1}|=\sum_{y=0}^{\gamma -2} |T_{y,\gamma -1}|$.  It
     is sufficient, then, to compute $\sum_{y=0}^{\gamma -2}
     |T_{y,\gamma -1}|$.
     
     Suppose $\{t_l\}_{l=0}^{s} \in T_{s-1,\gamma -1}$.  To this 
     sequence, we can associate a nonnegative
     planar walk of length $\gamma -1$ with $s+1$ vertices.  The walk begins at the
     origin, and steps are
     given by the
     vectors $(1,1)$ and $(1,-1)$.  The vertices are the origin, the
     endpoint of the walk, and the points where the
     direction of the walk changes from up to down or from down to up.  We make
     the association in the following way:  let $t_i$ be the $x$-coordinate
     of the $i$th vertex.  It is clear that this walk is nonnegative, as the
     condition defining membership in $T_{s-1,\gamma -1}$ guarantees nonnegativity of
     the associated walk.  Conversely, the $x$-coordinates of
     the $s+1$ vertices in a nonnegative planar walk of length $\gamma -1$ form a
     sequence in $T_{s-1,\gamma -1}$.  Therefore, this association
     gives a bijective correspondence between sequences in
     $\cup_{y=0}^{\gamma -2} T_{y,\gamma -1}$ and nonnegative planar walks of
     length $\gamma -1$.  It is a fact (see, for example,~\cite{Feller})
     that the number of nonnegative planar walks of length $\gamma -1$ is
     given by ${\gamma -1\choose \lfloor \frac{\gamma -1}{2} \rfloor }$.  This
     means that $\sum_{y=0}^{\gamma -2}
     |T_{y,\gamma -1}|={\gamma -1\choose \lfloor \frac{\gamma -1}{2} \rfloor }$.
     Consequently, $|L'_{\gamma}|={\gamma -1\choose \lfloor \frac{\gamma -1}{2} \rfloor
     }$.

     It remains to show that, in fact, a field of order bigger than 
     $N_{\gamma}$ elements is large enough so that
     a superregular $\gamma\times \gamma$ matrix can exist over it.  This
     may 
     be easily
     seen in the following way:  the determinant of each submatrix in
     $S_l$, $l\in \{ 1,\ldots ,\gamma \}$, is linear in $x_l$.  $N_l$ increases with
     $l$.  Thus, if we work over a field of
     order bigger than $N_{\gamma}$ elements, it is
     possible to successively substitute nonzero field elements for 
     $x_1,x_2,\ldots
     ,x_{\gamma}$ in such a way that the 
     determinant of each
     submatrix in $S_l$, $l\in \{ 1,\ldots ,\gamma \}$, is nonzero.  This completes
     the proof.  
\end{proof}

Using computer searches, we were able to determine the exact minimum
field size for small values of $\gamma$.  These may be seen in Table
1.  We observe that the upper bound $N_{\gamma}+1$ appears to grow
much more quickly than necessary:
\begin{table}[h]
  \begin{center}
      \begin{tabular}{|l|c|r|}
        \hline
        $\gamma$ & Minimum Field Size Required & Upper Bound ($N_{\gamma}$ +1) \\
        \hline
        3 & 3 & 3 \\
        4 & 5 & 5 \\
        5 & 7 & 11 \\
        6 & 11 & 27 \\
        7 & 17 & 77 \\
        8 & 31 & 233 \\
        9 & 59 & 751 \\
        10 & $\leq 127$ & 2495 \\
        \hline
     \end{tabular}
     \caption{Comparison of Actual Required Field Size and $N_{\gamma}+1$}
  \end{center}
\end{table}

\noindent It is still an open problem as to how this bound may be refined. 
Computer searches lead us to make the following conjecture; if true, it would offer a
significant improvement to the bound given here:
\begin{conjecture}\label{ConjBound}
For $\gamma \geq 5$, there exists a $\gamma\times \gamma$ superregular matrix over the field
$\F_{2^{\gamma-2}}$.
\end{conjecture}

\section{For Future Research:  Finding a Construction}
At this point, little is understood about how to construct
superregular matrices.  The problem of finding constructions appears
to be a very hard one.  One must find a way of guaranteeing that all
proper submatrices with any number of zeroes above the diagonal have a nonzero determinant and do so with additional
constraints coming from the Toeplitz structure.  In~\cite{ToepArray}, a method is given for
constructing, for any prime number $p$, a triangular Toeplitz array of
dimension $p$ having the property that all full square submatrices
have a nonzero determinant.  Unfortunately, there is no way of
extending this construction to the much more general situation we
consider here.

In~\cite{gl03r}, the following result is proven:
\begin{theo}\label{TotPos}      
Let $X$ be the $\gamma\times \gamma$ matrix given by
      \begin{equation*}
        X:=\left[\begin{array}{cccccc}
        1&0&\cdots&\cdots&\cdots&0\\
        1&1&\ddots&&&\vdots\\
        0&1&1&\ddots&&\vdots\\
        \vdots&\ddots&\ddots&\ddots &\ddots&\vdots\\
        \vdots&&\ddots&1&1&0\\
        0&\cdots&\cdots&0&1&1\\
        \end{array}\right].
      \end{equation*}
            Then, 
      \begin{equation*}
        X^{\gamma -1}=\left[\begin{array}{cccccc}
        1&0&\cdots&\cdots&\cdots&0\\
        {\gamma -1\choose 1} &1&\ddots&&&\vdots\\
        {\gamma -1\choose  2} &{\gamma -1\choose 1}&1&\ddots&&\vdots\\
        \vdots&\vdots&\ddots&\ddots &\ddots&\vdots\\
        {\gamma -1\choose \gamma -2}&\gamma -1\choose {\gamma -3}&\cdots&{\gamma
        -1\choose 1}&1&0\\ 
        1&\gamma -1\choose {\gamma -2}&\cdots&\cdots&{\gamma -1\choose 1}&1
        \end{array}\right],
      \end{equation*}
      where ${\gamma\choose \omega }$=$\frac{\gamma !}{\omega !(\gamma -\omega )!}$,
      is totally positive over the real numbers.  Thus, for a sufficiently
      large prime number $p$, taking the entries of this matrix modulo $p$ gives
      a superregular matrix.
\end{theo}
This result gives a construction insofar as one knows that, modulo a large enough
prime number, the matrix $X^{\gamma -1}$ is superregular.  It is not clear,
however, how one may give a good bound as to how large $p$ must be for a given
$\gamma$.
 
\noindent

\bibliographystyle{Bib}
\bibliography{SR2}

\end{document}